\documentclass[twocolumn,showpacs]{revtex4}

\usepackage{amssymb}
\usepackage{graphicx}
\setkeys{Gin}{width=0.9\columnwidth}   

\begin{document}

\title{Hydrodynamical instabilities in an expanding 
quark gluon plasma}
\author{C. E. Aguiar}
\email{carlos@if.ufrj.br}
\author{E. S. Fraga}
\email{fraga@if.ufrj.br}
\author{T. Kodama}
\email{tkodama@if.ufrj.br}
\affiliation{Instituto de F{\'\i}sica, Universidade 
Federal do Rio de Janeiro \\
Caixa Postal~68528, Rio de Janeiro, 21941-972, RJ, Brazil}

\begin{abstract}
We study the mechanism responsible for the onset of instabilities in a
chiral phase transition at nonzero temperature and baryon chemical 
potential. As a low-energy effective model, we consider 
an expanding relativistic plasma of quarks coupled to a chiral field, 
and obtain a phenomenological {\it chiral hydrodynamics} from a 
variational principle. Studying the dispersion relation for small
fluctuations around equilibrium, we identify the role played by 
chiral waves and pressure waves in the generation of instabilities. 
We show that pressure modes become unstable earlier than chiral 
modes.
\end{abstract}

\maketitle

\date{\today}



\section{Introduction}

It is commonly believed that QCD at very high temperatures and/or densities
allows for a new phase of strongly interacting matter, the quark-gluon
plasma (QGP). Compelling lattice QCD results point in that direction 
\cite{Karsch:2001vs}, and experiments in 
ultra-relativistic heavy-ion collisions 
\cite{rhic} under way at BNL-RHIC \cite{QM2004} and planned for CERN-LHC
attempt to glimpse at this elusive state of matter that was presumably
present in the early universe \cite{cosmo}.

Depending on the nature of the QCD phase transition, the hadronization
process of the expanding QGP generated in a high-energy heavy-ion collision
may proceed in a number of different ways \cite{review}. 
In the analysis of phase conversion of strongly interacting 
matter from a chirally symmetric to a chirally broken state in 
the phase diagram of QCD, a mechanism of supercooling is 
usually implied \cite{QM2004,cosmo}. The kinetics of domain 
formation and growth can then be described by the the mechanisms 
of nucleation or spinodal decomposition, depending on the 
degree of supercooling \cite{review}. An expanding plasma will 
probe, in principle, different temperatures and experiment both 
mechanisms. In the case of the early universe, 
it is well known that the time scales for the expansion during the 
QCD transition are so slow that phase conversion is driven 
by the nucleation of bubbles of true vacuum inside the metastable 
phase, as the universe reaches temperatures below a critical 
value \cite{cosmo}. On the other hand, the QGP 
presumably created in experiments in ultra-relativistic heavy-ion 
collisions expands at a pace several orders of magnitude faster 
than the primordial universe \cite{QM2004,rhic},
and might simply bypass nucleation, entering the domain of spinodal 
decomposition \cite{sudden,Scavenius:1999zc,Scavenius:2001bb,Shukla:2001xv,
explosive,polyakov-explosive}. 
Some results from CERN-SPS and BNL-RHIC suggest what has been called 
\textit{sudden hadronization} \cite{sudden} or \textit{explosive behavior} 
\cite{polyakov-explosive,explosive}. From the theoretical side, 
this phenomenon was recently associated with rapid changes in the 
effective potential of QCD near the
critical temperature, such as predicted, for instance, by the 
Polyakov loop model \cite{polyakov,polyakov-explosive}, 
followed by spinodal decomposition \cite{review,explosive}. 
Clearly, an understanding of the interplay between the
typical space and time scales 
of the expanding plasma is welcome. Some attempts in this
direction can be found in 
Refs.~\cite{quarks-chiral,ove,Scavenius:1999zc,Scavenius:2001bb,
paech,bravina,Shukla:2001xv,finite1,finite2,randrup,FK}.

In this paper, we discuss the mechanism responsible for the onset of
instabilities in an expanding plasma. As a phenomenological model to mimic
the case of the QGP, we use a relativistic plasma of quarks 
coupled to a chiral field. Although we derive a phenomenological 
\textit{chiral hydrodynamics} from a variational principle, 
we do not focus on the
numerical solution of the resulting hydrodynamic transport equations.
Rather, we consider the role played by chiral waves and pressure waves in
the generation of unstable modes. We show that \textit{mechanical}
instabilities set in earlier than one would expect from the analysis of the
thermodynamic potential decoupled from hydrodynamical modes. Therefore, the
spinodal lines are shifted from the values one obtains studying only the
chiral degrees of freedom, modifying the phase diagram.

It has been argued \cite{critical-point,critical2,Scavenius:2000qd} that the
first-order transition line which starts at the $T=0$ axis of the
baryon chemical potential ($\mu$) vs. temperature ($T$) phase diagram 
for QCD should come to an end at a critical point $(\mu _{E},T_{E})$. 
Beyond this endpoint, for $0<\mu <\mu _{E}$, the QCD transition 
should become a smooth crossover, where 
spinodal instabilities do not occur. Results from lattice
simulations at finite $\mu$ also point in this direction \cite{fodor,allton}. 
Here, we consider two illustrative scenarios for 
the temperature ($T$) vs. baryon chemical potential ($\mu$) phase 
diagram in which a first order line starts at a point 
$(\mu,T)=(\mu_c,0)$ and ends in a critical point, $(\mu_E,T_E)$, 
with $\mu_E<\mu_c$, beyond which, for $0<\mu<\mu_E$, the phase 
transition becomes a smooth crossover. 

To model the mechanism of chiral symmetry breaking present in QCD, we adopt
a simple low-energy effective chiral model: the linear $\sigma$-model
coupled to quarks \cite{gellmann}, which in turn comprise the hydrodynamic
degrees of freedom of the system. Similar approaches, relying on low-energy
effective models for QCD and making use of a number of techniques to treat
the expanding plasma, can be found in the literature 
\cite{quarks-chiral,ove,Scavenius:1999zc,Scavenius:2001bb,paech}. 
The gas of quarks provides a thermal bath in which the long-wavelength
modes of the chiral field evolve. The latter plays the role of an order
parameter in a Landau-Ginzburg approach to the description of the chiral
phase transition \cite{Scavenius:2001bb,paech}.

The paper is organized as follows. In section II, we present our 
effective field theory model. In section III, we derive what we 
call phenomenological {\it chiral hydrodynamics} from a variational 
formulation of relativistic hydrodynamics. In section IV, we 
study the onset of instabilities. In section V, we discuss our results. 


\section{The effective model}

Let us consider a chiral field $\phi=(\sigma,\vec{\pi})$, where $\sigma$ is
a scalar field and $\pi^i$ are pseudoscalar fields playing the role of the
pions, coupled to two flavors of quarks according to the Lagrangian: 
\begin{equation}
\mathcal{L} = \overline{q}[i\gamma^{\mu}\partial _{\mu} + \mu_q\gamma^0 -
W(\phi)]q + \frac{1}{2}\partial_{\mu}\phi \partial^{\mu}\phi - V(\phi)\; .
\label{lagrangian}
\end{equation}
Here $q$ is the constituent-quark field $q=(u,d)$ and $\mu_q=\mu/3$ is the
quark chemical potential. The interaction between the quarks and the chiral
field is given by 
\begin{equation}
W(\phi) = g \left(\sigma + i\gamma_5 \vec{\tau}\cdot \vec{\pi}\right) \; ,
\end{equation}
and 
\begin{equation}
V(\phi)= \frac{\lambda^2}{4} 
\left(\sigma^2 + \vec{\pi}^2 -v^2\right)^2 - h_q\sigma
\end{equation}
is the self-interaction potential for $\phi$. The parameters above are
chosen such that chiral $SU_{L}(2) \otimes SU_{R}(2)$ symmetry is
spontaneously broken in the vacuum. The vacuum expectation values of the
condensates are 
$\langle\sigma\rangle =\mathit{f}_{\pi}$ and $\langle\vec{\pi}\rangle =0$, 
where $\mathit{f}_{\pi}=93$~MeV is the pion decay constant.
The explicit symmetry breaking term is due to the finite current-quark
masses and is determined by the PCAC relation, giving 
$h_q=f_{\pi}m_{\pi}^{2}$, where $m_{\pi}=138$~MeV is the pion mass. This
yields $v^{2}=f^{2}_{\pi}-{m^{2}_{\pi}}/{\lambda ^{2}}$. The value of 
$\lambda^2 = 20$ leads to a $\sigma$-mass, 
$m^2_\sigma=2\lambda^{2}f^{2}_{\pi}+m^{2}_{\pi}$, equal to 600~MeV. 
In mean field theory,
the purely bosonic part of this Lagrangian exhibits a second-order phase
transition~\cite{Pisarski:1984ms} at $T_c=\sqrt{2}v$ if the explicit
symmetry breaking term, $h_q$, is dropped. For $h_q\ne 0$, the transition
becomes a smooth crossover from the restored to broken symmetry phases. 
For $g>0$, the finite-temperature one-loop effective potential also 
includes a contribution from the quark fermionic determinant 
\cite{Scavenius:2001bb,paech}.

In what follows, we treat the gas of quarks as a heat 
bath for the chiral field, with temperature $T$ and baryon-chemical 
potential $\mu$. Then, one can integrate over the fermionic degrees 
of freedom, obtaining an effective theory for the chiral field $\phi$. 
To compute thermodynamic quantities, one needs the partition function 
\begin{equation}
{\cal Z}= \int [{\cal D}\phi][{\cal D}q][{\cal D}\overline{q}] 
e^{-\int_0^{1/T} d\tau \int_{\cal V} d^3x {\cal L}_E}\quad, 
\label{partition}  
\end{equation}
where ${\cal L}_E$ is the Euclidean Lagrangian and ${\cal V}$ 
is the (infinite) volume of the plasma. Integrating over 
the fermions and using a classical approximation for the chiral 
field, we can write the thermodynamic potential as
\begin{equation}
\Omega(T,\mu,\phi) = V(\phi) - \frac{T}{\mathcal{V}} 
\ln\det\{[G_E^{-1}+W(\phi)] / T \} \; ,  
\label{TDpot}
\end{equation}
where $G_E$ is the fermionic Euclidean propagator satisfying 
\begin{equation}
\left[ \gamma^0\partial_{\tau} + i \vec{\gamma}\cdot\nabla 
-\mu\gamma^0  \right] G_E(\tau,\vec{x};\tau',\vec{x}')=
\delta^{(4)}(x-x')\; .
\label{propagator}
\end{equation}

{}From the thermodynamic potential, (\ref{TDpot}), one can obtain 
all the thermodynamic quantities of interest. The fermionic 
determinant that results from the functional integration over 
the quark fields can be calculated to one-loop order 
in the standard fashion \cite{Scavenius:2000qd,kapusta-book}. 
The effective Lagrangian for the chiral field, $\phi$, in the 
presence of the quark thermal bath is then
\begin{equation}
\mathcal{L}_{eff}^{(\phi)}=\frac{1}{2}\partial_{\mu}\phi\partial^{\mu} 
\phi-\Omega(T,\mu,\phi).
\label{Leff-fai} 
\end{equation}
%


\section{Chiral hydrodynamics}

Given the thermodynamic potential, one can derive the total pressure and
energy density, and obtain the conserved energy-momentum tensor,
$\mathcal{T}^{\mu\nu}$, for an expanding quark \textit{perfect fluid} 
by using standard methods of thermal field theory
(see, for instance, Ref. \cite{paech}). We prefer, instead, adopting an
alternative approach to obtain the hydrodynamic equations for our system:
the variational formulation \cite{Bailyn,variational}. This approach provides a
natural and unified way of merging chiral and fluid dynamics once the action 
of the system is specified. For a different treatment of the hydrodynamics of
nuclear matter in the chiral limit, see \cite{son}.

The variational formulation of relativistic hydrodynamics has been recently
applied to several physical systems, especially in the realms of astrophysics
and condensed matter. In the former it has been used, for instance, to
incorporate the effects of local turbulent motion in supernova explosion
mechanisms, and can prove to be useful in the analysis of the relativistic
motion of blast waves in models for gamma-ray bursts \cite{astro}. In soft
condensed matter, it has been used to study the bubble dynamics in
sonoluminescence experiments, in particular in deriving the relativistic
generalization of the Rayleigh-Plesset equation \cite{cond-mat}. In this
section, we briefly review how to obtain the usual equations of relativistic
hydrodynamics within this framework, and derive in detail the case in which 
we have a chiral field coupled to an expanding quark fluid.

We describe the state of the fluid in terms of the four-velocity $u^{\mu}(x)$,
the proper baryon density, $n(x)$, and the proper entropy density, $s(x)$. In
what follows we assume that $n$ and $s$ are conserved, \textit{i.e.}:
\begin{eqnarray}
\partial_{\mu}(nu^{\mu})&=&0\; , \nonumber\\
\partial_{\mu}(su^{\mu})&=&0\; .
\label{cons_ns}
\end{eqnarray}

The action that yields the hydrodynamic equations is given 
by \cite{Bailyn,variational}
\begin{equation}
S^{(fluid)}\equiv\int d^{4}x\left[  -\epsilon(n,s)\right]  \;,
\label{S_hydro} 
\end{equation}
where $\epsilon(n,s)$ is the proper energy density, from which
temperature and chemical potential are obtained via the 
usual thermodynamic relations, $T=\partial\epsilon(n,s)/\partial s$ 
and $\mu=\partial\epsilon(n,s)/\partial n$. The equation for the 
hydrodynamic motion of the fluid is obtained by imposing the 
variational principle with respect to $n$, $s$ and $u^{\mu}$, under 
the constraints (\ref{cons_ns}) and the normalization condition 
\begin{equation}
u^{\mu}u_{\mu}=1 \; .
\label{Co-3} 
\end{equation}
These constraints can be incorporated in the variational principle in 
terms of Lagrangian multipliers, $\lambda(x)$, $\zeta(x)$ and $w(x)$, 
so that one imposes:
\begin{eqnarray}
\delta\int d^{4}x\left[\right.  &-&\epsilon(n,s)+
\lambda\partial_{\mu}(nu^{\mu})+
\zeta\partial_{\mu}(su^{\mu}) \nonumber \\
&+&\left. \frac{1}{2}w\left(u^{\mu}u_{\mu}-1\right)
\right] =0,
\label{Va-0}
\end{eqnarray}
Equivalently, the fluid dynamics is given by the effective Lagrangian, 
$\mathcal{L}_{eff}^{\left(fluid\right)}=\mathcal{L}_{eff}^{\left(fluid\right)}
\left(n,s,u^{\mu},\lambda,\zeta,w\right)$, 
\begin{eqnarray}
\mathcal{L}_{eff}^{\left(fluid\right)}&=& 
-\epsilon(n,s)-nu^{\mu}\partial_{\mu}\lambda \nonumber \\
&-& su^{\mu}\partial_{\mu}\zeta+
\frac{1}{2}w\left(  u^{\mu}u_{\mu}-1\right) \; .
\label{L_eff} 
\end{eqnarray}
Now the variables $n$, $s$, $u^{\mu}$, $\lambda$, $\zeta$, 
and $w$ are independent.

Variations with respect to $n$, $s$ and $u^{\mu}$ yield
\begin{eqnarray}
-\mu-u^{\mu}\partial_{\mu}\lambda &=&0 \; , \nonumber\\
-T-u^{\mu}\partial_{\mu}\zeta &=&0 \; , \nonumber \\
-n\partial_{\mu}\lambda-s\partial_{\mu}\zeta+wu_{\mu}  
&=&0 \; , 
\label{Va-3} 
\end{eqnarray}
while variations with respect to $\lambda$, $\zeta$ and $w$ give simply 
the constraints (\ref{cons_ns}) and (\ref{Co-3}). From (\ref{Co-3}) 
and (\ref{Va-3}), one obtains
\begin{eqnarray}
w  &=&n\mu+Ts \nonumber\\
&=& \varepsilon+p \; ,
\label{enthalpy} 
\end{eqnarray}
where $p$ is the pressure, and identifies $w$ as  
the enthalpy density. Using the Gibbs-Duhem relation, 
$dp=sdT+nd\mu$, and the property 
$u^{\nu}(\partial_{\mu}u_{\nu})=0$, it is 
straightforward to obtain
\begin{eqnarray}
\partial^{\nu}\left(  wu_{\mu}u_{\nu}\right)   &=&n\partial_{\mu} 
\mu+s\partial_{\mu}T+\left(  \partial_{\mu}u_{\nu}\right)  \xi u^{\nu
}\nonumber\\
&=&\partial_{\mu}p+u^{\nu}\left(  \partial_{\mu}u_{\nu}\right)
\xi\nonumber\\
&=&\partial_{\mu}p \; ,
\label{div-2} 
\end{eqnarray}
which can be rewritten in the standard form 
\begin{equation}
\partial^{\nu}T_{\mu\nu}=0,\label{divTmunu}
\end{equation}
where 
\begin{eqnarray}
T_{\mu\nu} &=&wu_{\mu}u_{\nu}-g_{\mu\nu}p\nonumber\\
&=&\left(  p+\varepsilon\right)  u_{\mu}u_{\nu}-g_{\mu\nu}p\label{Tmunu-1}
\end{eqnarray}
is the usual energy-momentum tensor of the fluid.

It is important to notice that the effective Lagrangian, 
(\ref{L_eff}), evaluated in the proper comoving frame of 
the fluid, 
\begin{equation}
\mathcal{L}_{eff}^{\left(  fluid\right)  } 
=-\epsilon(n,s)+\mu n+Ts=p \; ,
\label{L_eff_prop}
\end{equation}
is nothing but \textit{minus} the thermodynamic potential of the system.
This fact will be useful to couple the chiral degrees of freedom to the
quark fluid motion.

The procedure outlined above for the derivation of relativistic 
hydrodynamics presents a number of nice features. First, it can 
be easily generalized to include other degrees of freedom such 
as the chiral field. Secondly, although we have implicitly used a 
Minkowski metric, the generalization to the case of curved 
space-times is straightforward. This will be discussed later, 
when we consider applications to cosmology. Furthermore, from a 
practical point of view, once the variational approach is established 
one can use this method to obtain the optimal parameters for a given 
family of trial solutions.

Let us now describe how to couple a chiral field to the fluid
within the framework of the variational principle. The effective 
Lagrangian for the chiral field, (\ref{Leff-fai}), should be 
interpreted as the proper value of the Lagrangian where the 
thermalized quark fluid is at rest. Therefore, Eq. (\ref{L_eff_prop}) 
suggests that the part corresponding to the quark pressure should be 
replaced by expression (\ref{L_eff}) in order to reconstruct the 
fluid motion of thermalized quarks. Apart from the constraints, 
we have the following action for the coupled system of the chiral 
field, $\phi$, and the thermalized quark fluid motion:
\begin{equation}
S^{\left(  \phi+fluid\right)  }=
\int d^{4}x\left\{  \frac{1}{2}(\partial_{\mu
}\phi)(\partial^{\mu}\phi)-\epsilon(n,s,\phi)\right\}  \; .
\end{equation}
Here, the energy density, $\epsilon$, is related to the thermodynamic
potential, $\Omega$, through the Legendre
transformation $\epsilon=\Omega+Ts+\mu n$.
Since $\Omega=-p$, this is nothing but the general thermodynamic 
relation $\epsilon=-p+Ts+\mu n$. Notice that now the energy density 
depends on the field variable $\phi$ through
$\Omega=\Omega\left(T,\mu,\phi\right)$ and, correspondingly, the second
thermodynamic law should read
\begin{equation}
d\epsilon=Tds+\mu dn+Rd\phi \; .
\label{thermo-2} 
\end{equation}
The corresponding Gibbs-Duhem relation becomes 
\begin{equation}
dp=sdT+nd\mu-Rd\phi \; ,
\label{Gibbs-Duhem-2} 
\end{equation}
where we have defined a new quantity $R$ (with four
components $R_{i}=\partial\Omega/\partial\phi_{i}$): 
\begin{eqnarray}
R  &\equiv&\left(\frac{\partial\epsilon(n,s,\phi)}
{\partial\phi}\right)_{n,s} \nonumber \\
&=&\left(  \frac{\partial\Omega(T,\mu,\phi)}
{\partial\phi}\right)_{T,\mu} \; .
\end{eqnarray}
The second line corresponds to Maxwell's relation and comes from
Eq. (\ref{Gibbs-Duhem-2}). $R$ can be written as
\begin{equation}
R=\frac{\partial V(\phi)}{\partial\phi}+g\phi\rho(T,\mu,\phi) \; ,
\end{equation}
where
\begin{equation}
\rho=\nu_{q}\int\frac{d^{3}k}{(2\pi)^{3}}\frac{1/E_{k}(\phi)}{e^{[E_{k} 
(\phi)-\mu_{q}]/T}+1}+(\mu_{q}\rightarrow-\mu_{q})
\end{equation}
is a scalar density for quarks. Here $\nu_{q}=12$ stands for the
color-spin-isospin degeneracy of the quarks, $E_{k}(\phi)=(\vec{k}^{2} 
+m_{q}^{2}(\phi))^{1/2}$, and $m_{q}(\phi)=(g^{2}\phi^{2})^{1/2}=g(\sigma
^{2}+\vec{\pi}^{2})^{1/2}$ plays the role of an effective mass for the quarks.

Thus, our coupled system is described by an effective Lagrangian 
\begin{eqnarray}
\mathcal{L}_{eff}^{\left(\phi+fluid\right)}&=&
\frac{1}{2}(\partial_{\mu}\phi)(\partial^{\mu}\phi)
-\epsilon(n,s,\phi)-nu^{\mu}\partial_{\mu}\lambda \nonumber \\
&-& su^{\mu}\partial_{\mu}\zeta+\frac{1}{2}w
\left(u^{\mu}u_{\mu}-1\right) \; .
\label{L-coupled} 
\end{eqnarray}

The variation with respect to $\phi$ gives the equation of motion
for the chiral field, 
\begin{equation}
\Box\phi=-R \; 
\label{FieldEqM} 
\end{equation}
The variation procedures with respect to the fluid variables are almost
exactly the same as before, except for the use of the new Gibbs-Duhem 
relation, (\ref{Gibbs-Duhem-2}). Thus, Eq. (\ref{div-2}) is modified to 
\begin{equation}
\partial^{\nu}\left(  wu_{\mu}u_{\nu}\right)  =\partial_{\mu}p+R\partial_{\mu
}\phi,\label{EqM-fluid} 
\end{equation}
or, in the standard form, 
\begin{equation}
\partial^{\nu}T_{\nu\mu}=R\partial_{\mu}\phi \; ,
\label{divTmunu2} 
\end{equation}
where $T_{\mu\nu}$ is given by Eq. (\ref{Tmunu-1}) and is not the 
total energy momentum tensor of the system, 
$\mathcal{T}^{\mu\nu}=T^{\mu\nu}-\frac{1} 
{2}\partial_{\alpha}\phi\partial^{\alpha}\phi g^{\mu\nu}+\partial^{\mu} 
\phi\partial^{\nu}\phi$, which is of course conserved.

Note that Eq. (\ref{divTmunu2}) contains terms from the pure contribution of
the field $\phi,$ such as $V\left(  \phi\right)  $ and $\partial
V/\partial\phi$, both in $T_{\mu\nu}$ and in $R$. However, as we can see from
the equivalent equation, (\ref{EqM-fluid}), some of these terms cancel out. To
see more clearly the coupling between the fluid equation and the
chiral field $\phi$, it may be convenient to identify the fluid contribution of
the pressure as
\begin{equation}
p^{\left(  fluid\right)  }=\frac{T}{\mathcal{V}}\ln\det[(G_{E}^{-1} 
+W(\phi))/T]\;,
\end{equation}
and the energy-momentum tensor of the fluid, 
\begin{equation}
T_{\mu\nu}^{\left(  fluid\right)  }=wu_{\mu}u_{\nu}-p^{\left(  fluid\right)
}g_{\mu\nu}.\label{Tmunu_f} 
\end{equation}
Then the hydrodynamical motion is written as
\begin{equation}
\partial^{\mu}T_{\mu\nu}^{\left(  fluid\right)  }-\rho(\phi)m_{q}\partial
_{\nu}m_{q}(\phi)=0\;.\label{conserv} 
\end{equation}
%


\section{Stability analysis}

A numerical analysis of the system of differential equations consisting of 
(\ref{conserv}) and the equations of motion for $\sigma$ and $\vec{\pi}$ 
was performed, for instance, in Refs. \cite{ove,paech}. Here, we refrain 
from this approach and, instead, consider the behavior of small oscillations 
around equilibrium to study the onset of instabilities.

In what follows, we expand around equilibrium keeping terms up to first 
order in the perturbation, {\it i.e.}:
\begin{eqnarray}
n(x) &=& n_{eq} + n_1(x) \; , \nonumber\\
s(x) &=& s_{eq} + s_1(x) \; , \nonumber\\
\phi(x) &=& \phi_{eq} + \phi_1(x) \; , \nonumber\\
u^{\nu}(x) &=& u^{\nu}_{eq} + u^{\nu}_1(x) \; .
\end{eqnarray}
Since the normalization of the four-velocity implies 
$u_1^{\nu}=(0,\vec{v}_1)$, and $R_{eq}=0$, the conservation laws 
(\ref{cons_ns}) yield
\begin{eqnarray}
\frac{\partial n_1}{\partial t}+n_{eq} \nabla\cdot\vec{v}_1=0\; ,\nonumber\\
\frac{\partial s_1}{\partial t}+s_{eq} \nabla\cdot\vec{v}_1=0\; ,
\label{conservation2}
\end{eqnarray}
and the chiral field equation of motion becomes
\begin{equation}
(\Box + m^2_{\phi})\phi=-\left(\frac{\partial R}{\partial n}\right)_{eq} n_1 
-\left(\frac{\partial R}{\partial s}\right)_{eq} s_1 \; ,
\label{wave}
\end{equation}
where we have defined the (diagonal) chiral mass matrix 
\begin{equation}
m^2_{\phi}\equiv 
\left[\frac{\partial^2\epsilon(n,s,\phi)}{\partial\phi^2}\right]_{eq}\; .
\end{equation}

The time component of 
Euler's equation, (\ref{EqM-fluid}), is automatically satisfied given the 
conservation laws above, and its spatial components yield
\begin{equation}
w_{eq}\frac{\partial\vec{v}_1}{\partial t}=-\nabla p_1 \; .
\label{euler2}
\end{equation}

For plane waves of the form $\psi_1(x)=\tilde{\psi}_1 e^{-iKx}$, where 
$K^{\mu}=(\omega,\vec{k})$ and $\psi_1$ stands for $n_1$, $s_1$, 
$\phi_1$ or $\vec{v}_1$, we can rewrite the wave equation, (\ref{wave}), 
and the spatial component of Euler's equation, (\ref{euler2}), as 
(dropping the tildes)
\begin{eqnarray}
(\omega ^{2}-{k}^{2}-m_{\pi }^{2}) ~\vec{\pi}_{1} &=&0\;,  
\label{pi_mode}
\\
(\omega ^{2}-{k}^{2}-m_{\sigma }^{2}) ~{\sigma }_{1} &=&\frac{k}{\omega}
w_{eq}R^{\prime }{v}_{1}\;,  
\label{sig_v} \\
(\omega ^{2}-p^{\prime }{k}^{2}) ~{v}_{1} &=&
\omega kR^{\prime }{\sigma }_{1}\;,  
\label{v_sig}
\end{eqnarray}%
where we have defined
\begin{equation}
p'\equiv 
\left[\frac{\partial p(\epsilon,n/s,\phi)}{\partial\epsilon}\right]_{eq}\; ,
\end{equation}
\begin{equation}
R'\equiv 
\left[\frac{\partial R_{\sigma}(\epsilon,n/s,\phi)}{\partial\epsilon} 
\right]_{eq}\; ,
\end{equation}
evaluated at constant $(s/n)$ and $\phi$, 
and used the conservation laws (\ref{conservation2}) in the form 
(dropping the tildes)
\begin{equation}
\frac{{n}_1}{n_{eq}}=\frac{{s}_1}{s_{eq}}=
\frac{k_1 v_1}{\omega}\; .
\end{equation}

One can see that pions decouple from the hydrodynamic 
modes, to first order, and play no major role in the following 
analysis. In this approximation, the interplay between chiral 
modes and pressure modes happen entirely in the sigma sector 
of the chiral model. The dispersion relation then reads
\begin{equation}
(\omega^2-p'{k}^2)(\omega^2-{k}^2-m^2_{\sigma})=
w_{eq} R'^2 {k}^2 \; .
\label{dispersion}
\end{equation}
For long wavelength fluctuations, we can approximate the roots 
of (\ref{dispersion}) by
\begin{equation}
\omega_p^2/k^2 = \left( p' -\frac{w_{eq}R'^2}{m^2_{\sigma}}\right) +
{\cal O}({k}^2)\; ,
\label{omega-p}
\end{equation}
\begin{equation}
\omega_{\sigma}^2=m^2_{\sigma}+{\cal O}({k}^2)\; .
\end{equation}

These waves can be identified as pressure modes of frequency $\omega_p$,
moving with the speed of sound
\begin{equation}
v_{sound} = \left(p' -\frac{w_{eq}R'^2}{m^2_{\sigma}}\right)^{1/2}\; ,
\label{vsound}
\end{equation}
and chiral modes with frequency $\omega_{\sigma}$ and mass 
\begin{equation}
m_{\sigma}^2 = \left[\frac{\partial^2\epsilon(n,s,\sigma)}{\partial\sigma^2}\right]_{eq}\; .
\label{msigma}
\end{equation}

The onset of instabilities takes place when 
$\omega^2<0$ \cite{review}, 
{\it i.e.}, either for $v_{sound}^2<0$ or for $m^2_{\sigma}<0$. 
We see from Eq.~(\ref{vsound}) that pressure modes become 
unstable before chiral modes do.

To study the onset of instabilities within the phase diagram of the 
low-energy effective model for QCD described previously, we consider 
two illustrative cases. 
The first corresponds to a coupling between quarks and the 
chiral field small enough for the transition at 
$\mu=0$ to be a smooth crossover. We use $g=3.3$, which
yields a constituent quark mass of $307~$MeV, about $1/3$ of 
the nucleon mass.
In this case there is a line of first-order transitions
starting at $T=0$, $\mu_c=920$~MeV, and ending 
at a critical point $E$ with $T_E=98$~MeV and 
$\mu_E=630$~MeV.
In the second scenario we take a larger coupling, $g=5.5$, 
so that even at $\mu =0$ there is a first-order phase transition, 
in this case at $T_c\approx 123$~MeV 
\cite{Scavenius:1999zc,Scavenius:2001bb,paech}. 


\section{Discussion of results}

\begin{figure}[hbt]
\begin{center}
\includegraphics{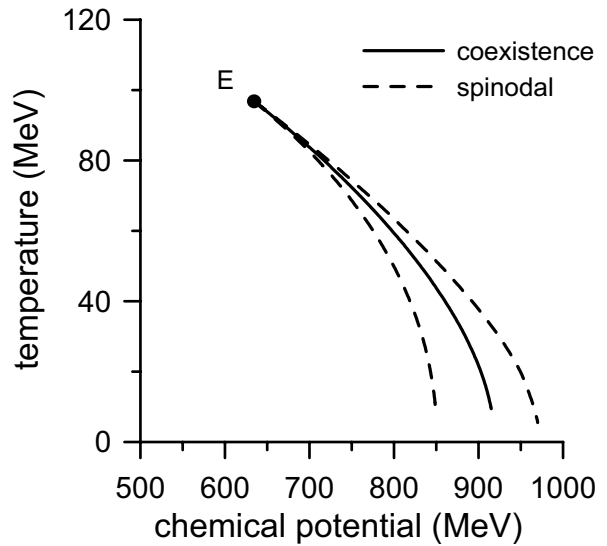}
\end{center}
\caption{Phase diagram for the effective model in the $(\mu,T)$-plane, 
for $g=3.3$. The coexistence line ending at the critical point $E$ is
represented by the full curve. 
The spinodal lines ($m_{\sigma}^2=0$) are shown as dashed curves.}
\label{muT-3.3}
\end{figure}

Let us now discuss the onset of instabilities in the
phase diagram of our effective model. 
In Fig.~\ref{muT-3.3} we plot the phase diagram in 
the $(\mu ,T)$-plane for the couplings $g=3.3$. 
The coexistence line is represented by the solid curve. 
The lines where pressure waves become unstable ($v_{sound}=0$) 
are shown as dashed curves. They correspond to the supercooling 
and superheating spinodals.

\begin{figure}
\begin{center}
\includegraphics{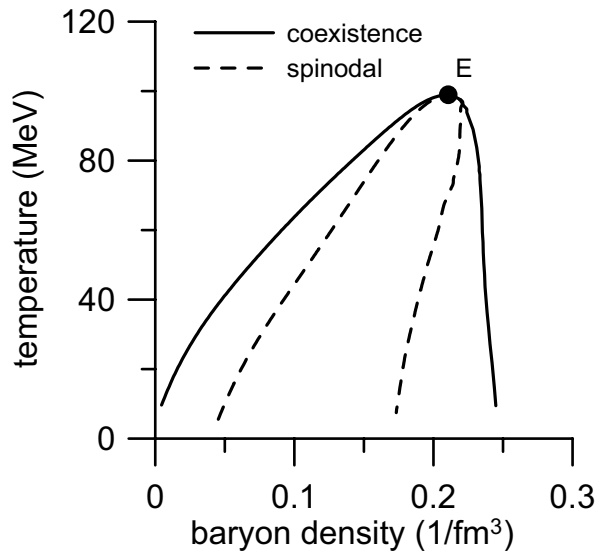}
\caption{Phase diagram for the effective model in the $(n,T)$ plane, 
for $g=3.3$. Line conventions are the same as in Fig.~1.  
Inside the full curve the system is mechanically unstable.}
\label{nT-3.3}
\end{center}
\end{figure}

The phase diagram in the $(n,T)$-plane is shown in Fig.~\ref{nT-3.3}.
The phase border of the coexistence region is represented by the solid
line and dashed lines stand for the spinodals. The sector between the lines 
on the right of the critical point $E$ corresponds to supercooled 
states in the chirally symmetric phase. Superheated states correspond to
the area on the left. The domain inside the dashed 
lines corresponds to mechanically unstable states which undergo
spinodal decomposition. 

\begin{figure}
\begin{center}
\includegraphics{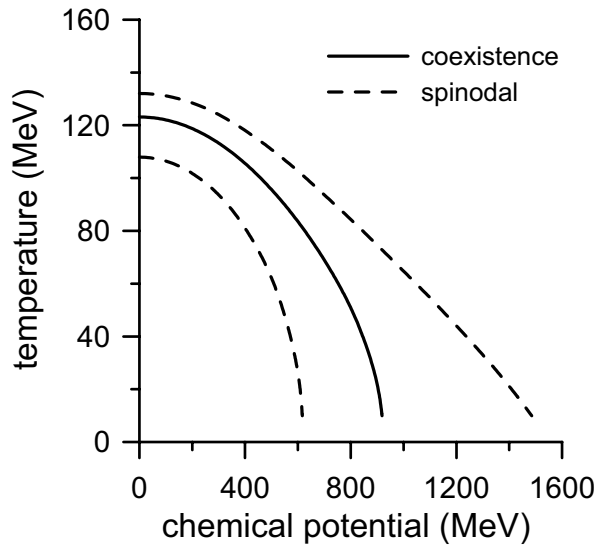}
\caption{The same as Fig. \ref{muT-3.3} for $g=5.5$.}
\label{muT-5.5}
\end{center}
\end{figure}

\begin{figure}
\begin{center}
\includegraphics{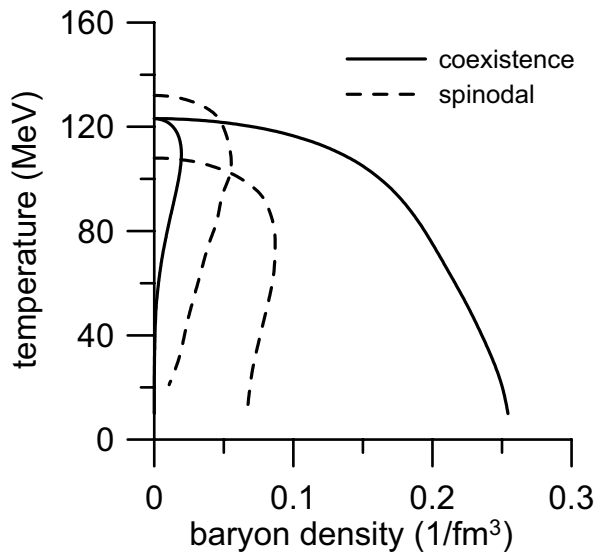}
\caption{The same as Fig. \ref{nT-3.3} for $g=5.5$.}
\label{nT-5.5}
\end{center}
\end{figure}

Results for the second choice of the coupling constant, $g=5.5$,
are shown in Figs.~\ref{muT-5.5} and \ref{nT-5.5}. As we have 
mentioned, the difference of this case with respect to the 
first is that the coexistence line now reaches $\mu=0$. 
Because of this, the spinodals
cross each other in the $(n,T)$ plane, and part of the phase 
coexistence region can be occupied by both supercooled and 
superheated states. Below the crossing no metastable states
exist.


We have shown that the process of domain formation and 
growth in the phase conversion of strongly 
interacting matter from a chirally symmetric to a chirally 
broken state in an expanding QGP is dominated by the 
exponential increase of hydrodynamical sound-like modes,
while chiral-like modes remain stable. We have mapped the 
boundaries of the unstable region for two illustrative cases, 
determining how much supercooling (or superheating) is necessary 
for the onset of spinodal instabilities. 

In the case of relativistic heavy ion reactions, the short time scale 
of expansion and effects due to the finite size of the system 
could change significantly our results. In order to 
address these points one should perform extensive numerical simulations 
to study the development of instabilities in different scenarios. 
Results in this direction will be presented elsewhere.


\begin{acknowledgments}
We thank A. Dumitru, K. Paech, D. Rischke, J. Schaffner-Bielich and
H. St{\"o}cker for fruitful discussions. 
This work was partially supported by CAPES, CNPq, FAPERJ and FUJB/UFRJ.
\end{acknowledgments}



\end{document}